\documentclass[twocolumn,showpacs,preprintnumbers,amsmath,amssymb]{revtex4}

\usepackage[dvips]{graphicx}
\usepackage{epsfig}

\begin{document}

\preprint{ }

\title{ Equations of Motion with Multiple Proper Time: A New Interpretation of Spin }  
\author{\normalsize Xiaodong Chen}
\email{xiaodong.chen@gmail.com} 

\date{\today}

\begin{abstract} 
  The purpose of this paper is to show that: when a single particle moving under
  3-proper time (three-dimensional time), the trajectories of a classical particle
 are equivalent to a quantum field with spin. 
Three-proper time models are built for spinless particle,
particles with integer spin and half-integer spin respectively. 
The models recreate the same physical behavior as quantum field
theory of free particles--- by using pure classical 
methods with three proper time. A new interpretation of spin is given. 
It provides us more evident that it is possible to interpret quantum physics by 
using multiple dimensional time. In the last part of this paper, 
Bose-Einstein statistics and Fermi-Dirac statistics are derived 
under classical method.

\end{abstract}

\maketitle

\section{Introduction} \label{INTRO}

There are three major properties of quantum physics 
which are different from classical physics:

1) Non-local property of single quantum particle: a single particle can stay in 
different places at the same time. Non-local property is the most important part which 
quantum physics is distinguish itself from classical physics. Without non-local property, the 
wave of single particle becomes an oscillator, which can be found in classical physics.

2) Statistical effect in the measurement of single particle. 

3) Spin and related statistics: Bose-Einstein statistics and Fermi-Dirac statistics.

In previous paper \cite{chen1} we demonstrate that: if a classical particle moves under
three proper time (three dimensional time), it will give the same non-local property of 
quantum particle and the same statistical effect of measurement as quantum particle. 
The properties 1) and 2) of quantum particle 
that we mentioned above can be derived from classical particle under three proper
time model. For instance, Fig1. draw the trajectory of single free particle
under two independent proper time $\tau$ and $\sigma$ on $x_0-x_i$ plane in Riemann Space.

\begin{figure}[t]
\begin{center}
\leavevmode
\hbox{%
\epsfxsize=2.5in
\epsffile{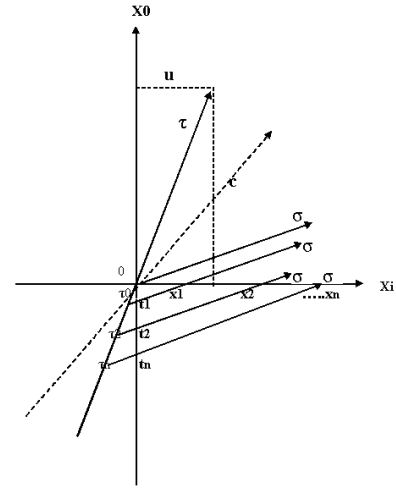}}
\caption{World line $\tau$ and world line $\sigma$ on $X_0-X_i$ plane in Riemann space.
At each points of world line $\tau$, the particle can also move along world line $\sigma$ 
by second proper time $\sigma$. The
slope of $\tau$ is u/c. Each world lines $\sigma$ parallel to each other with slope 
$v/c = c/u$. In Riemann space, $\tau$ and $\sigma$ are orthogonal to each other.
From above fig, we can see that, when particle moving with two proper time:
At $t=0$, the single particle will stay at many positions:
$x_1, x_2, ... x_n$ with different values of $\tau$ and $\sigma$ 
$(\tau_1,\sigma_1)..(\tau_n,\sigma_n)$; Also the particle
will stay at $x=0$ at different time: $t_1, t_2, .. t_n$ with
different values of $\tau, sigma$; where
$x_n = h/mu$ and $t_n = h/mc^2$ which are de Broglie wavelength 
and period. }
\end{center}
\label{fig:Fig1}
\end{figure}

In Fig1. a single particle's trajectory is determined by two proper time ($\tau$ and $\sigma$)
where world line $\tau$ and $\sigma$ are orthogonal to each other. 
At each points on world line $\tau$, particle will also move along world line $\sigma$. As
the result: At $t=0$, the single particle will stay at many positions:
$x_1, x_2, ... x_n$ with different values of $\tau$ and $\sigma$ 
$(\tau_1,\sigma_1)..(\tau_n,\sigma_n)$. Also the single particle
will stay at $x=0$ at different time: $t_1, t_2, .. t_n$ with
different values of $\tau, sigma$; where
$x_n = h/mu$ and $t_n = h/mc^2$ which are de Broglie wavelength 
and period. By adding periodic conditions for $\tau$ and $\sigma$, Fig1. becomes
de Broglie plane wave for quantum particle. Briefly speaking, in paper \cite{chen1},
we built 3 proper time model for free particle as below:

1)Along world line $\tau$, free particle moves the same as classical particle 
with classical energy and momentum. 

2)Free particle can move along world line $\sigma$ and $\phi$(where $\sigma$ and $\phi$ are second and third
proper time respectively) while first proper time $\tau$ unchanged. 
World lines of $\sigma$ and $\phi$ are also straight lines. 
Relations between $\sigma$ and 2nd time dimension $x_4$, 
$\phi$ and third time dimension $x_5$ are 
$x_4 = e^{i\sigma} = e^{\frac{im_0 \tau}{\hbar}}$, $x_5 = e^{i\phi} =
e^{\frac{-im_0 \tau}{\hbar}}$, 
These equations come from the geometry of $x_4$ and $x_5$, which are loops in complex plane.

The statistical effect of measurement for single particle is derived from the nature character
of time: the apparatus can meet a particle at a spatial location if and only if all three values of 
three dimensional time are equals between particle and apparatus: $t_i^{apparatus} = t_{i}^{particle} $.
But because we only have knowledge of one dimensional time, we don't know how to synchronize 
other two time dimensions, the results of measurement become statistical. (see paper \cite{chen1} 
for detail).

This paper will give more clear picture of 3-proper time model than previous paper \cite{chen1}. 
We will extend above 3-proper time model to include particle with spin.
We will derive spin, Bose-Einstein statistics and Fermi-Dirac 
statistics from classical physics under three proper time. 
In section \ref{ZeroSpin}, 3-proper 
time model is introduced for spinless particle. In section \ref{OneSpin}, 3-proper
 time model for Boson with spin $> 0$ is derived. In section \ref{HalfSpin}, 3-proper
 time for Fermion is discussed. In section \ref{Statistic}, we derived 
 Bose-Einstein statistics and Fermi-Dirac statistics for Boson and Fermion respectively.
In last section, we will give the interpretation of what is spin.
 The purpose of this paper is to show that: when a single particle moving under 
 3-proper time, the trajectories of a classical particle 
 are equivalent to a quantum field with spin. 

\section{3-proper time models for free spinless single particle} \label{ZeroSpin}

We need clarify some terminologies which will be used in this paper.
we will use both words: three dimensional time and three proper time. 
Proper time is the same meaning as it is in relativity:
Proper time is a special affine parameter of space-time;
 Each proper time has one related world line, i.e. each proper time corresponding to one individual movement.
Time dimension means the time value in
a chosen time coordinates system. 
We will use word : ``the direction of speed of particle'' which
means the spatial moving direction of the particle with proper time $\tau$.
``World line $\tau$'' means the trajectory of particle's motion by proper time $\tau$. Similarly we will use
``World line $\sigma$'' and ``World line $\phi$ ''.

Through out this paper, we keep some reasonable assumptions:

1)Particle's world lines by different proper time are orthogonal to each other.

2)All three proper time are independent. 

3)We implement cylinder condition on 2nd and 3rd proper time 
$\sigma$, and $\phi$: Each of them is angle of a loop with value from $0$ to $2\pi$.

For spinless particle, the 3-proper time model is:

1) World line $\tau$ is the same as the world line in relativity; The velocity of particle in world line $\tau$ is $u$, which is
the same as velocity in relativity, it is also the group velocity of de Broglie wave.

2) The projection of world line $\sigma$ in 4-dimensional time-space
$(x_0,\vec{x})$ is orthogonal to world line $\tau$; The velocity of particle on world line $\sigma$ is $v$,
which is the phase velocity of de Broglie wave.

3) The projection of world line $\phi$ in 4-dimensional time-space
$(x_0,\vec{x})$ is coincident with world line $\tau$.

Considering particle under rest reference frame, 
velocity on world line $\tau$ is zero: $u_i = 0$. In this case, phase velocity of de Broglie wave
v becomes infinite since $ v= \omega \lambda = c^2/u $.  To understand this, considering in
rest frame, proper time $\tau = t$, world line $\tau$ becomes $X_0$ axis in Fig1., 
world line $\sigma$ becomes $X_i$ axis in Fig1.;
Particle does not move with $\tau$, but it 
still moves with second proper time $\sigma$. The
definition of velocity with $\sigma$ is $ v_i = c dx_i/dx_0$ where c is speed of light,
since this motion does not change first dimensional time t:
$dx_0 = 0$,  so $v_i$ becomes infinite. To obtain the equation of motion under $\sigma$, 
we need know the relationship between $\sigma$ and 4-dimensional space-time coordinates. 
In previous paper \cite{chen1}, world line $\sigma$ can go infinite since plane wave can be
everywhere in the space. In reality, the inital position of particle is always under certain boundary condition.
Now we localize the particle's wave-length up to one wave-length.
Let second time dimension $x_4 = \sigma$, the spatial direction
of moving by $\sigma$ is $x_1$, and choose equation
of motion in rest frame be:
\begin{equation}
\sigma = \pi \cos{\frac{m_0 c}{\hbar} x_1} \; \; \; \; 0 \leq x < \frac{h}{m_0 c}   
\label{EqMotion1}
\end{equation}
and 
\begin{eqnarray}
x_1 = x_5    \\
\label{EqMotion1_1}
x_j = 0 \; \; \; \; j=0,2,3   \\
\label{EqMotion1_2}
\sigma = x_4              \\
\label{EqMotion1_3}
\end{eqnarray}
where $m_0$ is rest mass of particle, c is speed of light, h is planck constant; $x_4$ is 2nd time dimension;
$x_5$ is 3rd time dimension. 
Equation (\ref{EqMotion1}) tells us that in the rest reference frame,
particle moves like an oscillator with proper time $\sigma$.
The equation contains $m_0$, so we can understand that the motion of oscillating 
comes from the rest energy: $m_0 c^2$. 
Back to general reference frame with velocity $u = u_1 > 0$, by using Lorentz transformation, we turn back to Fig1.,
where particle is still non-localized within one wave length. 
Using Lorenze transformation, and noticing that $v_1 = c^2/u_1$, Equation (\ref{EqMotion1}) becomes:
\begin{equation}
\sigma = \pi \cos{\frac{im_0 c}{\hbar} \frac{(v_1 t - x_{1})}{\sqrt{1-v_1 ^2/c^2}} } = 
	\pi \cos{\frac{1}{\hbar}(Et-p^1 x_1)}
\label{EqMotion2}
\end{equation}
where $x_1$ and t are coordinates in new reference frame. The imaginary number i is to keep
$\sigma$ being real number in Lorenze transformation.  
We can rewite equation (\ref{EqMotion2}) as
\begin{eqnarray}
\sigma = \frac{\pi}{2} ( a e^{\frac{-i}{\hbar}p_1 x_1}  + a^{*} e^{\frac{i}{\hbar}p_1 x_1 }) 
\label{EqMotion3_0}
\end{eqnarray}
where $a^{*}$ is complex conjugate of a, which is 
\begin{equation}
a = \frac{\pi}{2} e^{\frac{i}{\hbar} Et}
\label{Eq:a}
\end{equation}
Equation (\ref{EqMotion3_0}) is derived under the condition that the direction of motion is point to $x_1$. In general case, it can be
written as:
\begin{eqnarray}
\sigma = \frac{\pi}{2} ( a e^{\frac{-i}{\hbar}p^{i} x_i}  + a^{*} e^{\frac{i}{\hbar}p^{i} x_i }) 
\label{EqMotion3}
\end{eqnarray}
where i= 1,2,3.
In quantum field theory, $a$ in equation(\ref{EqMotion3}) is annihilation operator of scalar field, $a^{*}$ 
is creation operator. Here we see that $a^{*}$ means particle moves toward positive t 
direction, $a$ means particle moves toward negative t direction. 

Let the relation between second proper time $\sigma$ and second time dimension $x_4$ be:
\begin{equation}
d\sigma = \cos{\frac{1}{\hbar}(p^{\alpha} x_{\alpha} - m_0 x_5)} dx_4
\end{equation}
The difference between $\sigma$ and $x_4$ comes from the geometry of time-space. On world line $\sigma$, it becomes
equation(\ref{EqMotion1_3}).

The geometry of 6-dimensional time-space is:
\begin{equation}
ds^2 = dx_{\alpha}dx^{\alpha} +\psi^2 dx_4 dx^4 - dx_5 dx^5
\label{metric1}
\end{equation}
where $\psi$ can be chosen as
\begin{equation}
\psi = e^{\frac{i}{\hbar}(p^{\alpha} x_{\alpha} - m_0 x_5)}
\label{psi_1}
\end{equation}
or
\begin{equation}
\psi = \cos{\frac{1}{\hbar}(p^{\alpha} x_{\alpha} - m_0 x_5)}
\label{psi_2}
\end{equation}
The metric of 6-dimensional space-time is
\begin{equation}
\left( \hat{g}_{AB} \right) = \left( \begin{array}{cc}
   g_{\alpha\beta} \; \; \; \; \; \;  \; \; \; \; \; \; \\
  \; \; \; \; \; \; \; \; \psi \; \; \; \; \; \; \; \; \\
   \; \; \; \;  \; \; \; \; \; \; \; \; \; \; -1 \\ 
   \end{array} \right) 
\label{6dMetric_0}
\end{equation}
where metric elements $g_{\alpha\beta}$ is 4-dimensional metric. 
Substitute above metric into Einstein field equation \cite{chen2} \cite{chen3}:
\begin{equation}
\hat{G}_{AB} = \kappa \hat{T}_{AB} \; \; \; ,
\label{5dEFE1}
\end{equation}
We can derive Klein-Golden equation:
\begin{equation}
\partial_{\alpha} \partial^{\alpha} \psi + m_0^2 \psi= 0
\label{wavefunction}
\end{equation}
where $\alpha$ is 0..3. 
From above discussion, we see that non-local property and field equation
of quantum scalar field can be derived by classical particle moving under 
3 proper time. Look at ``time'' part of equation (\ref{metric1}):
\begin{equation}
ds^2_{t} = dx_0 - dx_5 dx^5 + \psi dx_4 dx^4
\end{equation}
Compare to geometry of sphere:
\begin{equation}
ds^2_{t} = dr^2 - d\theta^2- r^2\sin{\theta}^2 d\phi^2
\end{equation}
The 2nd and 3rd time dimension is similar to sphere with unit radius.

\section{3-proper time models for free Boson with spin one and spine $> 1$} \label{OneSpin}

In section \ref{ZeroSpin}, for spinless particle, the projection of world line $\phi$ on 4-dimensional 
space-time $(t, \vec{x})$ is coincident with world line $\tau$. In this section we
will see that, when the projections of world lines $\tau$, $\sigma$, $\phi$ on 4-dimensional space-time $(t, \vec{x})$
are separated, we get equations of quantum particle with integer spin $>1$. 

\subsection{3-proper time model for photon}

Since the rest mass of photon is zero, we 
can not use equation (\ref{EqMotion1}). Instead, the oscillation for photon
comes from electric energy and magnetic energy. For free photon with single frequency,
Electric field and magnetic field both are perpendicular to the direction of wave-vector $\vec k$.
Now we choose $x_1$ axis as direction of electric field 
$\vec E$, $x_2$ axis as direction of magnetic field $\vec B$, $x_3$ axis as
direction of wave-vector $\vec k$. First, we build 3-proper time model in rest frame
where Photon's speed with world line $\tau$ is zero.
In world line $\sigma$, we choose the equation of motion as:
\begin{eqnarray}
x_1 = \lambda (E_0)\arccos{\frac{\sigma}{\pi}} \\
x_3 = \lambda \arccos{\frac{\sigma}{\pi}} 
\label{eq:x3_0}
\end{eqnarray}
where k is wave-vector of photon, $\lambda (E_0)$ is coefficient dependent on magnitude of electric field $E_0$. So 
$\sigma$ can be derived from either $x_1$ or $x_3$. From equation(\ref{eq:x3_0}), we have:
\begin{eqnarray}
\sigma = \pi \cos{k x_3} 
\label{Eq:electric}
\end{eqnarray} 
Similarly for $\phi$:
\begin{eqnarray}
x_2 = \lambda (B_0)\arccos{\frac{\phi}{\pi}} \\
x_3 = \lambda \arccos{\frac{\sigma}{\pi}} 
\end{eqnarray}
\begin{equation}
\phi = \pi \cos{kx_3}
\label{Eq:magnetic}
\end{equation}
Where $\lambda (B_0)$ is a variable the dependent on magnitude of magnetic field $B_0$. 

In regular reference frame, photon moves at direction $x_3$ with speed c. The equation for $\sigma$ is:
\begin{equation}
\sigma = \pi \cos{(\omega t - k x_3)} 
\label{EqMotionPhoton1}
\end{equation}
and the equations of motions for $\phi$ is:
\begin{equation}
\phi = \pi \cos{(\omega t - k x_3)}
\label{EqMotionPhoton2}
\end{equation}
We know that electric field E and 
magnetic field B of photon perpendicular to wave-vector $\vec{k}$. Let
\begin{equation}
\vec{E} = \alpha_e \sigma \vec{e_1} 
\label{EqMotionPhoton3}
\end{equation}
\begin{equation}
\vec{B} = \alpha_b \phi \vec{e_2} 
\label{EqMotionPhoton4}
\end{equation}
where $\alpha_{e(b)}$ is constant; $\vec{e_1}$, $\vec{e_2}$ are unit vector point to $x_1$ and $x_2$ axis.
Now we derived photon plane wave by using classical motion of single photon under 3-proper time. 

\subsection{3-proper time model for free massive Boson }

As we have seen in this paper: besides the classical motion on proper time $\tau$;
For spinless particle, it has extra oscillation by proper time $\sigma$, 
the oscillation is caused by rest energy; for photon, it has two
extra oscillations by $\sigma$ and $\phi$ which are  
perpendicular to the direction of wave-vector, they are caused by electric field E and 
magnetic field B. Now for a massive Boson with spin one, it can have 
three separated oscillations. Also in previous sections, we assume that, in rest frame,
the world line of first proper time $\tau$ is only moving toward $x_0$ dirction, $t = \tau$.
That is not a required condition. In general, we can assume that: in rest frame,
world line $\tau$ oscillates around t, while the average effect of motion is the same as classical proper time
$\tau$.

For free particle with spin one, let:

1) Three independent proper time be: $\tau$, $\sigma$, $\phi$.

2) Each world line perpendicular to each other  in 3-dimensional space. 

3) We localized each world lines so that, they are all cosine functions.

In rest frame,  Let the equations of motion for world line $\sigma$ be 
\begin{eqnarray}
x_1 = \lambda (V_{10})\arccos{\frac{\sigma}{\pi}} \\
\label{eqV11}
x_3 = \frac{\hbar}{m_0 c} \arccos{\frac{\sigma}{\pi}} 
\label{eqV12}
\end{eqnarray}
and from equation(\ref{eqV12}), we can represent $\sigma$ by $x_3$:
\begin{equation}
\sigma = \pi \cos{(\frac{m_0 c}{\hbar}x3 )}   
\label{Eq:Motion7}
\end{equation}
Where $\lambda (V_{10})$ is a variable the dependent on magnitude of vector field $V_{10}$. 
The equations of motion for world line $\phi$ be:
\begin{eqnarray}
x_2 = \lambda (V_{20})\arccos{\frac{\phi}{\pi}} \\
\label{eqV21}
x_3 = \frac{\hbar}{m_0 c} \arccos{\frac{\phi}{\pi}} 
\label{eqV22}
\end{eqnarray}
Where $\lambda (V_{20})$ is a variable the dependent on magnitude of vector field $V_{20}$. 
And from equation(\ref{eqV22}), we can represent $\sigma$ by $x_3$:
\begin{equation}
\phi = \pi \cos{(\frac{m_0 c}{\hbar}x3 )}   
\label{Eq:Motion8}
\end{equation}
The equations of motion for world line $\theta$ be:
\begin{equation}
x_3 = \frac{\hbar}{m_0 c} \arccos{\frac{\tau}{\pi}} 
\label{eqV3}
\end{equation}
And:
\begin{equation}
\tau = \pi \cos{(\frac{m_0 c}{\hbar}x3 )}   
\label{Eq:Motion9}
\end{equation}

In the reference frame with speed of particle $u_3 > 0$ toward direction of $x_3$, 
by using Lorentz transformation:
we have:
\begin{equation}
\sigma = \pi \cos{\frac{1}{\hbar}(Et - p^i x_i)} 
\label{Eq:Motion10}
\end{equation}
\begin{eqnarray}
\phi = \pi \cos{\frac{1}{\hbar}(Et - p^i x_i)} \\
\label{Eq:Motion11}
\theta = \pi \cos{\frac{1}{\hbar}(Et - p^i x_i)} 
\label{Eq:Motion12}
\end{eqnarray}
When $m_0$ is zero, above equations
becomes the equations of photon. Equations(\ref{Eq:Motion10})(\ref{Eq:Motion11})(\ref{Eq:Motion12})
can also be written as:
\begin{equation}
\sigma = a_1 e^{\frac{i}{\hbar}p^i x_i } + a_1^{*} e^{\frac{-i}{\hbar} p^i x_i}
\label{Eq:Motion14}
\end{equation}
\begin{equation}
\phi = a_2 e^{\frac{i}{\hbar}( p^i x_i )} + a_2^{*} e^{\frac{-i}{\hbar}( p^i x_i)}
\label{Eq:Motion15}
\end{equation}
\begin{equation}
\tau = a_3 e^{\frac{i}{\hbar}( p^i x_i )} + a_3^{*} e^{\frac{-i}{\hbar}( p^i x_i)}
\label{Eq:Motion16}
\end{equation}
where $a_i^{*}$ is complex conjugate of a which is 
\begin{equation}
a_i = \frac{\pi}{2} e^{\frac{-i}{\hbar} Et}
\label{Eq:a2}
\end{equation}
$a_i$ is equivalent to annihilation operator of scalar field, $a^{*}$ is creation operator. The
model assume that vector $V_1$, $V_2$ perpendicular to the direction of speed of particle
\begin{equation}
\vec{V_1} = \alpha_e \sigma \vec{e_1}
\label{EqMotionPhoton5}
\end{equation}
\begin{equation}
\vec{V_2} = \alpha_b \phi \vec{e_2} 
\label{EqMotionPhoton6}
\end{equation}
where $\alpha_{e(b)}$ is constant.

We can also see that the components of world lines on $x_3$ axis (the direction of speed) gives wave functions for
vector field; the components of world lines on $x_1$ and $x_2$ axis gives the mangnitude of vector field.

\subsection{Lorentz Covariant Vector Fields}

The above equations(\ref{Eq:Motion14})(\ref{Eq:Motion15})(\ref{Eq:Motion16}) contains
three independent motions, they are not invariant under Lorentz transformation. To obtain Lorentz invariant
vector fields: using the new fields $A_\alpha$ ($\alpha = 0...3$), where 
\begin{eqnarray}
\nabla \hat{A_0} + \partial_{x_0} \partial \vec{\hat A}  = \sigma \vec{e_1}   \\
\label{Eq:Motion17}
\nabla \times \vec{\hat A} = \phi \vec{e_2} \\
\label{Eq:Motion18}
\partial_{x_5} \partial \vec{\hat A} = \theta \vec{e_3}
\end{eqnarray}
where $\hat{A}( \hat{A_0},\vec{\hat A})$ is 4-vector which is Lorentz invariant, $\vec{e_i}$ is unit vector point to
the same direction of $x_i$ axis. When $m_0$ is zero, 
then all $\partial_{x_4}$ becomes zero, above equations becomes the equations of photon.

In paper \cite{chen2} \cite{chen3}, I approved that by chosen time-space metric:
\begin{equation}
\left( \hat{g}_{AB} \right) = \left( \begin{array}{cc}
   g_{\alpha\beta} + \hat{A}_{\alpha} \hat{A}_{\beta} \; \; & \; \; 
       \hat{A}_{\alpha} \; \\
    1 \; \;                                & \; \; 
     \hat{A}_{\beta} \; \; \; \; \; \; \\
  \; \; \; \; \; \; \; \; \;  & \; \; \; \; \; \; \; \; \; \; \;  \; \;   -1  \end{array} \right) 
\label{6dMetric1S_m}
\end{equation}
where A, B equals 0...5. $\alpha$, $\beta$ equals 0...3. And put above metric into Einstein field 
equation(\ref{5dEFE1}), one can get the same quantum field equations of vector field:
\begin{equation}
\frac{1}{4}\hat{F}_{\alpha\beta} \hat{F}^{\alpha\beta} - \frac{1}{2} m_{0}^{2} \hat{A}_{\alpha} \hat{A}^{\alpha} = 0 
\label{vectorF}
\end{equation}
and Proca equation:
\begin{equation}
\partial_{\alpha} \hat{F}_{\alpha\beta} + m_0^{2} \hat{A}_{\beta} = 0 
\label{planePhoton}
\end{equation}
Here we obtain field equations for quantum vector field by using pure classical method
(general relativity equations). When $m_0 = 0$, above equations become Maxwell-equation for free photon in vacuum.

Under time-space metric, we can write world line as:
\begin{equation}
ds^2 = dx_\alpha dx^{\alpha} - dx_5 dx^5 +  (\hat{A}_{\alpha} dx^{\alpha} + dx_4)^2
\label{worldLine}
\end{equation}
where $\alpha$ is 0...3.  From equation(\ref{worldLine}) we can see that, if we choose 2nd time coordinate as
\begin{equation}
dx_{new}^{4} = \hat{A}_{\alpha} dx^{\alpha} + dx^4
\end{equation}
We got local flat time-space.

For general boson particle with spin $> 0$, we can get the field equations by using similar 
methods: i.e. first, create three independent motion vectors by three independent proper time. Then
convert them to 4-vectors. 

\subsection{Lorentz Covariant Vector Fields for general electro-magnetic fields}

When $m_0 = 0$, equation(\ref{vectorF}) is only satisfied for free plane wave photon. 
Generally, it is not satisfied by regular electro-magnetci fields: for instance, static electric field.
But static electric field is not particle. There are two matters in relativity: particle and curved space-time.
Static electric field can be treated as curved space-time. The 6-dimensional time-space metric for regular electro-magnetic fields is:
\begin{equation}
\left( \hat{g}_{AB} \right) = \left( \begin{array}{cc}
   g_{\alpha\beta} + A_{\alpha} A_{\beta} \; \; & \; \; 
       A_{\alpha} \; \\
    1 \; \;                                & \; \; 
     A_{\beta} \; \; \; \; \; \; \\
   A_{5} A_{\beta} \; & \; \; \; A_5  \; \; \; \;  \;   -1 + A_5 A_5  \end{array} \right) 
\label{6dMetric1S_elec}
\end{equation}
where A, B equals 0...5. $\alpha$, $\beta$ equals 0...3. 
put above metric into Einstein field equation(\ref{5dEFE1}), we get:
\begin{equation}
\frac{1}{2}F_{\alpha\beta} F^{\alpha\beta} - \partial_{\alpha} A_5 \partial^{\alpha} A^5 = 0 
\label{vectorF2}
\end{equation}
\begin{equation}
\partial_{\alpha} F_{\alpha\beta} = 0 
\label{planePhoton2}
\end{equation}
and equation
\begin{equation}
\partial_{\alpha} \partial^{\alpha} A_5 = 0 
\label{electroMagnetic}
\end{equation}
For static electric field, let 
\begin{equation}
A_5 = \frac{e}{r} 
\end{equation}
$A_5$ satisfied equations (\ref{vectorF2}) and (\ref{electroMagnetic}).

\section{3-proper time model for free single particle with half integer spin} \label{HalfSpin}

In section \ref{ZeroSpin} and \ref{OneSpin}, the trajectories of particle with
2nd and 3rd proper time are straight lines.
In this section, we will see that: when the trajectories of particle with 2nd and 3rd
proper time become circles, we will obtain equations of motion for particle with the half-integer spin.

For free particle with half spin, in rest frame (i.e. in the fram $t=\tau$), 
we build 3-proper time model for $\tau$, $\sigma$, $\phi$
as below:

1) Choosing a reference direction for the rotations by $\sigma$ and $\phi$. Let
trajectory of $\sigma$ parallel to $x_1 - x_2$ plane, $x_3$ perpendicular to world lines
$\sigma$ (this is equivalent to choose $x_3$ representation for spin).

2) Let trajectory of $\phi$ be a rotation from $x_3$ axis to $\vec{s}$ where $\vec{s}$ is
unit vector in $x_1 - x_2$ plane. 

The circles of rotation must be small. It is resonable to assume that the diameter of circle equals 
$h/m_0 c$ , so the radius of circle is 
\begin{equation}
r_{0} = \frac{h}{2m_0 c} 
\end{equation}
Let the equations of motion of $\sigma$ be:
\begin{eqnarray}
x_1 = r_{0}\cos{\frac{m_0 c}{\hbar}\sigma} \\
\label{Eq:halfSpin1}
x_2 = r_{0}\sin{\frac{m_0 c}{\hbar}\sigma}
\label{Eq:halfSpin2}
\end{eqnarray}
We can see that $x_1$ and $x_2$ has a phase difference $\pi /2$ .
Let the equations of motion for $\phi$ be:
\begin{eqnarray}
x_3 = r_{0}\cos{\frac{m_0 c}{4\hbar} \phi} \\
\label{Eq:halfSpin3}
x_s = r_{0}\sin{\frac{m_0 c}{4\hbar} \phi}
\label{Eq:halfSpin4}
\end{eqnarray}
let the equations of motion for $\tau$ be:
\begin{eqnarray}
x_0 = \tau (1 + \cos{(\frac{m_0 c}{\hbar} \tau)})  \\
x_5 = \tau \sin{(\frac{m_0 c}{\hbar} \tau)} 
\label{Eq:halfSpin5}
\end{eqnarray}
The motions of world lines $\sigma$ and $\phi$ are oriented, suppose that on world line
$\phi$ at point $\alpha_0$, particle's moving direction of world line $\sigma$ is perpendicular
to $\phi$ towards its right, and assume the direction of rotation towards positive $x_3$ direction. When it moves to the 
opposite point of $\alpha_0$ at sphere, its motion of world line $\sigma$ is still perpendicular to $\phi$ towards its right, 
but the direction of rotation towards negative $x_3$ direction which is opposite to previous motion. So, to build a
stable two-proper time motion on spherical surface, we have to limit all the motions on half sphere.
That's why we have factor $\frac{1}{4}$ inside equations (\ref{Eq:halfSpin3})
(\ref{Eq:halfSpin4})(\ref{Eq:halfSpin5}): $\phi$ is from 0 to $2\pi$, 1/4 factor to make the rotation from $x_3$ to $x_s$
with angel from 0 to $\pi /2$. 

For the reference frame with velocity of particle $u > 0$,
define one vector for each motion. For world line $\sigma$:
\begin{equation}
\vec{V_1} = (\vec{e_1} + i \vec{e_2}) e^{\frac{i}{\hbar}(Et - p^i x_i)} 
\label{Eq:halfSpin6}
\end{equation}
where i comes from the $\pi /2$ phase difference of $x_1$ and $x_2$ since world line $\sigma$ is a rotation.
For world line $\phi$:
\begin{equation}
\vec{V_2} = \vec{e_3}  e^{\frac{i}{\hbar} (Et - p^i x_i)} 
\label{Eq:halfSpin7}
\end{equation}
We ignored the contribution for $\vec{e_s}$ since the direction of $\vec{e_s}$ is from 0 to $2\pi$, the sum of contribution
will be zero.
For world line $\tau$, define:
\begin{equation}
\vec{V_3} =  \vec{e_{\tau}} + (\vec{e_{0}} + i\vec{e_{5}}) e^{\frac{i}{\hbar}(Et - p^i x_i)} 
\label{Eq:halfSpin8}
\end{equation}
where $\vec{e_{\tau}}$ is the direction of speed of particle. In rest frame, it is the same as $x_0$ axis.
In previous section, the components of each world lines on direction of speed of particle give wave-function.
Here we let:
\begin{eqnarray}
\psi_1 = \vec{V_1} \cdot \vec{e_{n}} = \sinh{\frac{\alpha}{2}} (\frac{u_1}{u} + i\frac{u_2}{u})e^{\frac{i}{\hbar}(Et - p^i x_i)}  \\
\label{halfWave1}
\psi_2 = \vec{V_2} \cdot \vec{e_{n}} = \sinh{\frac{\alpha}{2}} \frac{u_3}{u} e^{\frac{i}{\hbar}(Et - p^i x_i)}  \\
\label{halfWave2}
\psi_3 = \vec{V_3} \cdot \vec{e_{n}} = \cosh{\frac{\alpha}{2}} e^{\frac{i}{\hbar}(Et - p^i x_i)} + 1  
\label{halfWave3}
\end{eqnarray} 
where $u = \sqrt{u_1^2 + u_2^2 + u_3^2}$ is the speed of particle; $\vec{e_n}$ is the direction of $\hat{n}$ shown in Fig2.

\begin{figure}[t]
\begin{center}
\leavevmode
\hbox{%
\epsfxsize=2.5in
\epsffile{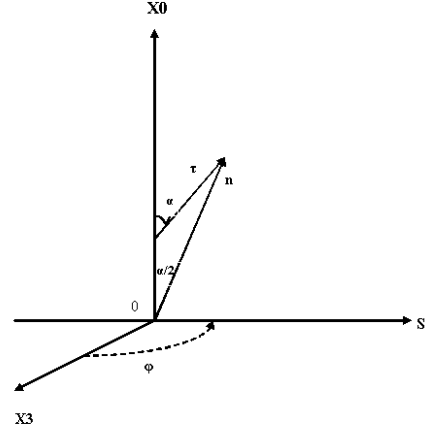}}
\caption{World lines in $x_0 - x_3 -x_s$ coordinates for Fermion.}
\end{center}
\label{fig:Fig2}
\end{figure}

\begin{equation}
\vec{e_n} = \vec{e_0} + \vec{e_{\tau}}
\label{Eq:en}
\end{equation}
and
\begin{eqnarray}
\cosh{\alpha} = \frac{1}{\sqrt{1-\frac{u^2}{c^2}}} \\
\sinh{\alpha} = \frac{\frac{u}{c}}{\sqrt{1-\frac{u^2}{c^2}} }
\end{eqnarray}
It is easy to see that:
\begin{eqnarray}
\psi_1 = \frac{p_1 + ip_2}{m_0}e^{\frac{i}{\hbar}(Et - p^i x_i)} = C_0 \psi_4^{D}  \\
\label{halfWave4}
\psi_2 = \frac{p_3}{m_0}e^{\frac{i}{\hbar}(Et - p^i x_i)}  = C_0 \psi_3^{D}  \\
\label{halfWave5}
\psi_3 = \frac{p_0}{m_0}e^{\frac{i}{\hbar}(Et - p^i x_i)} + 1 = C_0 \psi_0^{D} + 1
\label{halfWave6}
\end{eqnarray} 
where $\psi_0^{D}$,$\psi_2^{D}$,$\psi_3^{D}$ are three non-zero components of the solution of Dirac equation in $x_3$ representation,
with positive energy and spin. $C_0$ is normalization constant. The constant item 1 can be ignored. Therefore, $\psi_i$ corresponding
to the three non-zero components of Dirac wave-function.

Now we define 5-vector $\hat{K}$, the first 4 components are:
\begin{eqnarray}
\hat{K_0} = C\psi_3 e^{im_0 x_5}   \; \; \; \;
\hat{K_1} = -C\psi_1 e^{im_0 x_5}  \nonumber \\ 
\hat{K_2} = iC\psi_1 \phi_3 e^{im_0 x_5} \; \; \; \;
\hat{K_3} = -C\psi_2 e^{im_0 x_5}  
\label{HalfSpinVector}
\end{eqnarray}
where C is constant. If we modify equation(\ref{Eq:en}) as:
\begin{equation}
\vec{e_n} = \vec{e_0} + \vec{e_{\tau}} + \vec{e_5}
\label{Eq:en_2}
\end{equation}
Then
\begin{equation}
\psi_3 = \vec{V_3} \cdot \vec{e_{n}} = \cosh{\frac{\alpha}{2}}(1+i) e^{\frac{i}{\hbar}(Et - p^i x_i)} + 1  
\label{halfWave7}
\end{equation} 
The extra item $ie^{\frac{i}{\hbar}(Et - p^i x_i)}$ gives us 5th components of $\hat{K}$:
\begin{equation}
\hat{K_5} = -Ce^{\frac{i}{\hbar}(Et - p^i x_i- m_0 x_5)}   
\end{equation}

The geometry of 6-dimensional time-space for Fermion is \cite{chen3}: 
\begin{equation}
\left( \hat{g}_{AB} \right) = \left( \begin{array}{cc}
   g_{\alpha\beta} + \hat{K}_{\alpha} \hat{K}_{\beta} \; \; & \; \; 
      \hat{K}_{\alpha} \; \; \; \; \; \;  \; \;  \hat{K}_{\alpha} \hat{K}_{5} \\
    \hat{K}_{\beta} \; \;  & \; \; 
     1 \; \; \; \; \; \;  \; \;   \hat{K}_5  \\ 
   \hat{K}_{5} \hat{K}_{\beta} \; \; & \; \;  \hat{K}_5 \; \; \; \;  \; \;  -1+\hat{K}_{5} \hat{K}_{5}  \end{array} \right) \;
\label{6dMetricHalfS}
\end{equation}
Paper \cite{chen2} \cite{chen3} show that: put above $ \hat{g}_{AB} $ into Einstein field equation, We can get Dirac equation.
Under metric equation (\ref{6dMetricHalfS}), the interval $ds$ can be written as:
\begin{equation}
ds^2 = dx_\alpha dx^{\alpha} - dx_5 dx^5 +  (\hat{K}_{\alpha} dx^{\alpha}+ \hat{K}_5 dx^5 + dx^4)^2
\label{worldLine2}
\end{equation}
we can see that, if we choose 2nd local time coordinate $x_4$ as
\begin{equation}
dx^{4}_{new} = \hat{K}_{\alpha} dx^{\alpha} + \hat{K}_5 dx^5  + dx^4
\end{equation}
We obtain local flat time-space. I.e. Local inertia reference frame exists in this 6-dimesional time-space.

\section{Interpretation of Bose-Einstein Statistics and Fermi-Dirac Statistics } \label{Statistic}

Consider an event A: particle 1 and particle 2 interact each other at spatial point $O(x_1,x_2,x_3)$. In classical
4-dimensional time-space, event A can happen 
if and only if both particle meet at $O(x_1,x_2,x_3)$ at the same time t. In another words: their world lines across
each other at point $O(t,\vec{x})$. In 3+3 dimensional time-space, event A can happen if and only if the world lines of 
particle 1 and particle 2 across each other at $O(t, \vec{x}, x_4, x_5)$. If world lines of particle 1 and world lines of 
particle 2 parallel each other, then there is no interaction between them (Here we neglect the case of 3rd particle involving 
interaction).

\begin{figure}[t]
\begin{center}
\leavevmode
\hbox{%
\epsfxsize=2.5in
\epsffile{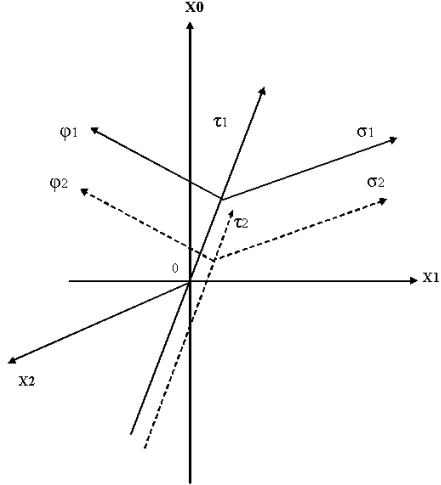}}
\caption{World lines $\tau$, $\sigma$ and $\phi$ of two Bosons paralles each other.}
\end{center}
\label{fig:Fig3}
\end{figure}

For Boson, Fig3. shows a distribution of world lines of particle 1 and particle 2, where their world lines $\tau$, 
$\sigma$ and $\phi$ parallel each other respectively, particle 2 always slightly ``later'' than particle 1. Because particle 
is ``point particle'', the size of particle is infinitesimal, then inside a spatial volume with the size of one wave-length, we
can put infinite particles inside where their world lines are parallel to each other. No interactions will be occurs inside this volume. We get 
Bose-Einstein condensation. Also because we can not synchronize the 2nd and 3rd time dimension, then we can not identify each
particle individually -- they are identical particles.

For Fermion, section \ref{HalfSpin} shows that, the world lines $\sigma$ and $\phi$ filled half spherical surface with diameter
equals $h/m_0 c$. In the spatial volume with the size of $h/m_0 c$, we can not find two parallel spherical without crossing each
other. But we can have another particle on the other half spherical surface with $x_1-x_2$ rotation point to negative $x_3$ direction.
That is, within the volume with the size of $h/m_0 c$, there can only have two non-interactive Fermions with rotations towards
opposite directions. This is Fermi-Dirac statistics.

Above interpretation is based on the equations for free Boson and Fermion. I believe that the basic concepts can be extended 
to cases of non-free particles.

\section{Summary and Discussions } \label{Discussions}

Now we are back to the initial topic of this paper: what is spin? As we have seen in previous sections 

1) For spinless particle, the projections  of world lines $\tau$, $\sigma$ and $\phi$ on 3-dimensional space 
are coincident on the same spatial straight lines.

2) For particle with integer spin $>0$, the projections of world lines $\tau$, $\sigma$ and $\phi$ on 3-dimensional space 
are three separate lines. Both $\sigma$ and $\phi$ have spatial components which perpendicular to spatial direction of
world line $\tau$.

3) For particle with half-integer spin, world lines $\sigma$ and $\phi$ become rotations on a spherical surface (like a real
spin). 

As the summary, the spin is inner 6-dimensional geometry properties of particle, it determines the motion of three-proper 
time of the particle.

This paper only study the case of free particle, it will be interesting to see the behavior of 3-proper time motions for interactive
particles. I believe that 3-proper time model will not only reproduce the physics of quantum field theory, but also give more
detail and more physics for quantum particle. To find the difference between 3-proper time model and quantum physics will be another
interesting topic.

Finally, I'd like to say that, there will be many different 3-proper time models. In future, there will be 3-proper time models
which may better fit into quantum physics than the model proposed by this paper. Again, the purpose of this paper is to show
that, 3-proper time model provides a possible way to interpret the physical effect of spin particles and basic quantum physics.

\end{document}